\def\kms{km s$^{-1}$}
\def\hi{H{\sc i}}
\def\hii{H{\sc ii}}

\def\mjyb{mJy beam$^{-1}$}
\def\jyb{Jy beam$^{-1}$}

\def\cmdos{cm$^{-2}$}
\def\cm{cm$^{-3}$}

\def\radec{RA,Dec.(J2000)}

\documentclass{aa} 

\usepackage{graphicx}
\usepackage{txfonts}
\usepackage{natbib}

\begin{document}

   \title{Molecular gas associated with IRAS 10361-5830}

           
   \author{M. M. Vazzano\inst{1,}\inst{2}, 
           C. E. Cappa\inst{1,}\inst{2}, 
           J. Vasquez\inst{1,}\inst{2}, 
           M. Rubio\inst{3} 
           \and 
           G. A. Romero\inst{2} }

   \institute{\inst{1} Instituto Argentino de Radioastronom\'ia, CONICET, CCT La Plata, C.C.5, 1894, Villa Elisa, Argentina \\
           \email{mvazzano@fcaglp.unlp.edu.ar}\\
           \inst{2} Facultad de Ciencias Astron\'omicas y Geof\'isicas, Universidad Nacional de la Plata, Paseo del Bosque s/n, 1900, La Plata, Argentina \\
           \inst{3} Departamento de Astronom\'ia, Universidad de Chile, Chile
            }

   \date{Received ...., 2013; accepted }

 
  \abstract
   {}
 {We analyze the distribution of the molecular gas and the dust in the molecular clump linked to \object{IRAS 10361-5830}, located in the environs of the bubble-shaped \hii\ region Gum 31 in the Carina region, with the aim of determining the main parameters of the associated material and investigating the evolutionary state of the young stellar objects identified there.}
   {Using the APEX telescope, we mapped the molecular emission in the J = 3-2 transition of three CO isotopologues, $^{12}$CO, $^{13}$CO and C$^{18}$O, over a $1\farcm 5 \times 1\farcm 5$ region around the IRAS position. We also observed the high density tracers CS and HCO$^{+}$ toward the source. The cold dust distribution was analyzed using submillimeter continuum data at 870 $\mu$m obtained with the APEX telescope. Complementary IR and radio data at different wavelengths were used to complete the study of the ISM.}
   {The molecular gas distribution reveals a cavity and a shell-like structure of $\sim$0.32 pc in radius centered at the position of the IRAS source, with some young stellar objects (YSOs) projected onto the cavity. The total molecular mass in the shell and the mean H$_2$volume density are $\sim$ 40 M$_{\odot}$ and $\sim$ (1-2)$\times 10^{3}$ cm$^{-3}$, respectively. The cold dust counterpart of the molecular shell has been detected in the far-IR at 870 $\mu$m and in $Herschel$ data at 350 $\mu$m. Weak extended emission at 24 $\mu$m from warm dust is projected onto the cavity, as well as weak radio continuum emission.}
   {A comparison of the distribution of cold and warm dust, and molecular and ionized gas allows us to conclude that a compact \hii\ region has developed in the molecular clump, indicating that this is an area of recent massive star formation. Probable exciting sources capable of creating the compact \hii\ region are investigated. The 2MASS source \object{10380461-5846233} (\object{MSX\,G286.3773-00.2563}) seems to be responsible for the formation of the \hii\ region.}


\maketitle
%

\section{Introduction}

The borders  of expanding \hii\ regions have proved to be excellent environs for the formation of dense molecular clumps where conditions for the triggering of new generations of stars are favoured (\citealt*{Elmegreen77}).
Dense molecular clumps and cores  can be studied through the line emission from low and high density tracers, like for example CS, HCO+, C$^{18}$O (e.g. \citealt{Dedes11}). The cold dust counterparts  of these dense regions, responsible for their high visual extinction,  can be detected by their continuum emission at submillimeter wavelengths (\citealt{Sanchez-Monje08}, \citealt{Beltran06}). 

High-density clumps where star formation is active can be identified by the presence of infrared point sources whose spectral energy distributions are typical of young stellar objects (YSOs), by OH- and/or  H$_{2}$O-maser emission, and by bipolar outflows detected in CO, SiO y HCO$^{+}$ lines (e.g. \citealt{Dionatos10}). 

In this paper we report on the results of a high angular resolution molecular line and dust continuum study of one of the high density clumps identified in the environs of the bubble-shaped \hii\ region \object{Gum\,31}. This \hii\ region is located in the Carina spiral arm and is excited by the open cluster \object{NGC\,3324}.
Distance estimates for \object{NGC\,3324} and Car\,OB1 are in the range 1.8-3.6 kpc (see \citealt{cappa08} [hereafter CNAV08] and \citealt {Ohlendorf13} for a discussion of the distance of the nebula). Following \citet{yonekura05} and \citet{Barnes11} we adopt a distance $d=2.5$ kpc, with an uncertainity of $\pm$0.5 kpc. 

\citet{yonekura05} and CNAV08 analyzed the molecular environment of \object{Gum\,31} using  C$^{18}$O(1-0) and $^{12}$CO(1-0) line data obtained with the NANTEN telescope with an angular resolution of 2\farcm 7. CNAV08 found an expanding molecular envelope encircling the ionized region detected in the velocity interval from --27.2 to --14 km s$^{-1}$ with ($7.6\pm3.4)\times10^4$ M$_{\odot}$ (at $\it d$ = 2.5 kpc) and an H$_2$ density of 230 cm$^{-3}$. The brighter CO(1-0) emission regions coincide with dense C$^{18}$O(1-0) clumps identified by \citet{yonekura05}. 

\begin{figure}
\includegraphics[width=0.5\textwidth]{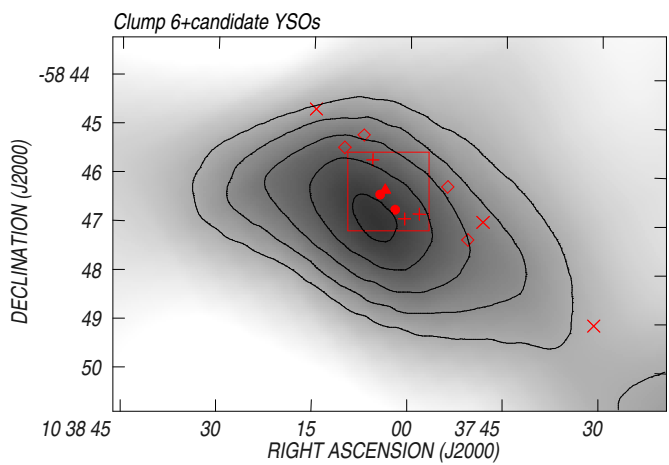}
\label{fig:grumo+fuentes}
\caption{$^{12}$CO(1-0) line emission distribution towards clump number 6 from the list \citet{yonekura05}, in \radec\ coordinates, extracted from CNAV08. Contour lines go from 4.0 to 8.0 K in steps of 1 K. The 4.0 K contour delineates approximately the clump as defined in C$^{18}$O. The greyscale goes from 0 to 10 K. The large square indicates the region observed in molecular lines with APEX(see Sect. 2.1). The triangle indicates the position of IRAS 10361-5530. The different symbols mark the location of candidate YSOs identified by CNAV08 and \citet{Ohlendorf13}: MSX sources (circles), 2MASS sources (crosses), Herschel sources (diamonds) and WISE sources (plus signs).}
\end{figure}

Here, we investigate the distribution of the molecular gas and cold dust in clump number 6 from the list of \citet{yonekura05}, named \object{BYF\,77} in the Census of High-and Medium-mass Protostars (ChaMP, \citealt{Barnes11}). Figure \ref{fig:grumo+fuentes} shows the $^{12}$CO(1-0) line emission distribution of clump 6 extracted from CNAV08 in contours and grayscale. Using CLUMPFIND algorithm, \citet{Barnes11} detected 4 cores in the HCO$^{+}$ data of \object{BYF\,77} obtained with MOPRA, with an angular resolution of 36\arcsec. \citet{yonekura05} found the highest density section of the clump using the H$^{13}$CO$^{+}$ line 1\farcm 6 to the west and south of the IRAS position. From their CO data they estimate a mean H$_{2}$ column density of 12.8$\times 10^{21}$ cm$^{-2}$, a H$_{2}$ mass of 3700 M$\odot$ and an H$_{2}$ volume density of 1600 cm$^{-3}$ adopting a radius of 1.8 pc for the clump and {\it d} = 2.5 kpc.

An infrared cluster spatially coincident with this clump was identified by \citeauthor{Dutra03} (2003, [DBS2008]\,128), while an X-ray clustering was identified by \citet{Preibisch14}. 

CNAV08 showed that this clump coincides with a number of YSOs detected in the IRAS, MSX, and 2MASS databases, indicating that star formation is active in this area. Its location on the molecular envelope bordering Gum 31 suggests that star formation in the clump was triggered by the expansion of the \hii\ region. This clump coincides with \object{IRAS\,10361-5830}, centered at \radec\ = (10$^{h}$38$^{m}$04.0$^{s}$, --58\degr 46\arcmin 17\farcs 8), and with the IR point like  sources \object{MSX\,G286.3579-00.2933}, \object{MSX\,G286.3747-00.2636}, \object{MSX\,G286.3773-00.2563}, \object{2MASS\,10375219-5847133}, and \object{2MASS\,10381461-5844416}. All of these were classified as candidate YSOs based on photometric criteria and are listed in Table 3 of CNAV08. Following Yamaguchi et al. (2001), the IR luminosity of \object{IRAS\,10361-5830} is 2$\times$10$^4$ L$_{\odot}$, suggesting massive star formation. 

An analysis aimed to characterize the protostellar and young stellar population of \object{NGC\,3324} and its environs was recently performed by \citet{Ohlendorf13},  who detected a dozen infrared point sources projected onto the dense clump using the Spitzer, WISE, and Herschel databases. They found some Class I objects and concluded that \object{MSX\,G286.3773-00.2563} (= J10380461-5846233) is a Class II object. The position of the candidate YSOs projected onto clump 6 identified by CNAV08 and \citet{Ohlendorf13} is also shown with different symbols in Fig. \ref{fig:grumo+fuentes}.

The characteristics of this dense clump make it an interesting object to investigate in more detail the molecular gas and dust distribution in the environs of the candidate YSOs, with the aim of analyzing the morphology and kinematics of the associated dense gas, and getting an additional insight on the evolutionary status of the inner sources.

\section{Observations}

\subsection{Molecular line data}

To accomplish  this study we surveyed a region of 90\arcsec $\times$90\arcsec\ centered on the IRAS position in the $^{12}$CO(3-2), $^{13}$CO(3-2), and C$^{18}$O(3-2) lines using APEX-2 (SHeFI) receiver (system temperature $T_{sys} \sim$  300 K) of the Atacama Pathfinder EXperiment (APEX) telescope, in December 2010. One single pointing was also observed towards the position of the IRAS source in the CS(7-6) and HCO$^+$(4-3) lines. The surveyed region is indicated in Fig. 1 by the large square. 
 
The frequency of the  observed lines, the integration time $\tau$ and the half-power beam-width (HPBW) are indicated in Table \ref{table:lineas}. The data were acquired with a FFT spectrometer, consisting of 4090 channels, with a total bandwidth of $\sim$ 800 \kms\ which provides a velocity resolution of  0.20 \kms\ . The observations were performed in the position switching mode with full sampling (i.e. with a separation of 10\arcsec). The off-source position free of  CO emission is located at \radec\ = (10$^h$38$^m$53.46$^s$, \hbox{--59\degr 18\arcmin 48\farcs 6)}.

Calibration was performed using the planet Saturn, RAFGL5254, and RAFGL4211 sources. Pointing was checked twice during observations using  X-TrA, the planet Venus, and VY-CMa. The intensity calibration has an uncertainty of  10\%.

The spectra were reduced using the Continuum and Line Analysis Single-dish Software (CLASS) of the Grenoble Image and Line Data Analysis Software (GILDAS)\footnote{http://www.iram.fr/IRAMFR/PDB/class/class.html}. A linear baseline fitting was applied to the data. The {\it rms} noise of the profiles after baseline subtraction and calibration is listed in Table \ref{table:lineas}. The observed line intensities are expressed as main-beam brightness-temperatures $T_{mb}$, by dividing the antenna temperature $T^{\ast}_A$  by the main-beam efficiency $\eta_{mb}$, equal to  0.82 for APEX-2 \citep{Vassilev08}. The Astronomical Image Processing System (AIPS) package and CLASS software were used to perform the analysis.

\subsection{Submillimeter continuum data}

We also mapped the submillimeter emission  in a field of 180\arcsec $\times$180\arcsec\ in size centered on the IRAS position with an angular resolution of 19\farcs 2 (HPBW), using the LArge Apex Bolometer Camera (LABOCA) \citep{siringoetal09} at 870 $\mu$m (345 GHz) operating with 295 pixels at the APEX telescope.
The field was observed during  1.9 hr in October 2011. The atmospheric opacity was measured every 1 hr with skydips. Atmospheric conditions were very good ($\tau_{zenith} \simeq$ 0.15). Focus was optimized on the planet Mars once during observations. The absolute calibration uncertainty is estimated to be 10\%. Data reduction was performed using The Comprehensive Reduction Utility for SHARC-2 software (CRUSH)\footnote{http://www.submm.caltech.edu/~sharc/crush/index.html}. The  noise level is in the range 15-20 \mjyb.\

\subsection{Complementary data}

 The millimeter and submillimeter data were complemented with Spitzer images at 3.6, 4.5, 5.8, and 8.0 $\mu$m from the Galactic Legacy Infrared Mid-Plane Survey Extraordinaire (GLIMPSE) \citep{benjaminetal03}, and images at  24 $\mu$m from the MIPS Inner Galactic Plane Survey (MIPSGAL) \citep{Carey09}. Also an image at 843 MHz from the Sydney University Molonglo Sky Survey (SUMSS, \citealt{mauchetal03}, synthesized beam =  $43\arcsec \times 43\arcsec$ cosec$(\delta)$) was used. 

\begin{table}
\centering
\begin{tabular}{cccccc}
\hline
{\it Line} & $\nu$ [GHz]  & $\tau$ [seg] & HPBW [\arcsec] & $\sigma$ [K] \\
\hline
  $^{12}$CO\,(3-2) & 345.79   & 5     &  18.9  & 0.40  \\
  $^{13}$CO\,(3-2) & 330.58   & 108   &  21.8  & 0.20  \\
  C$^{18}$O\,(3-2) & 329.33   & 108   &  21.8  & 0.65  \\
  CS\,(7-6)        & 342.88   & 300   &  18.2  & 0.21  \\
  HCO$^{+}$\,(4-3) & 356.73   & 600   &  17.5  & 0.03  \\
\hline
\end{tabular}
\caption{Observational parameters of the rotational lines: the frequency $\nu$, the integration time $\tau$, the HPBW, and the {\it rms} noise of each spectrum after baseline subtraction and calibration $\sigma$. }
\label{table:lineas}
\end{table}

\section{Results}

\subsection{Spectral analysis}

\begin{figure*}
\centering
\includegraphics[angle=270,width=0.9\textwidth]{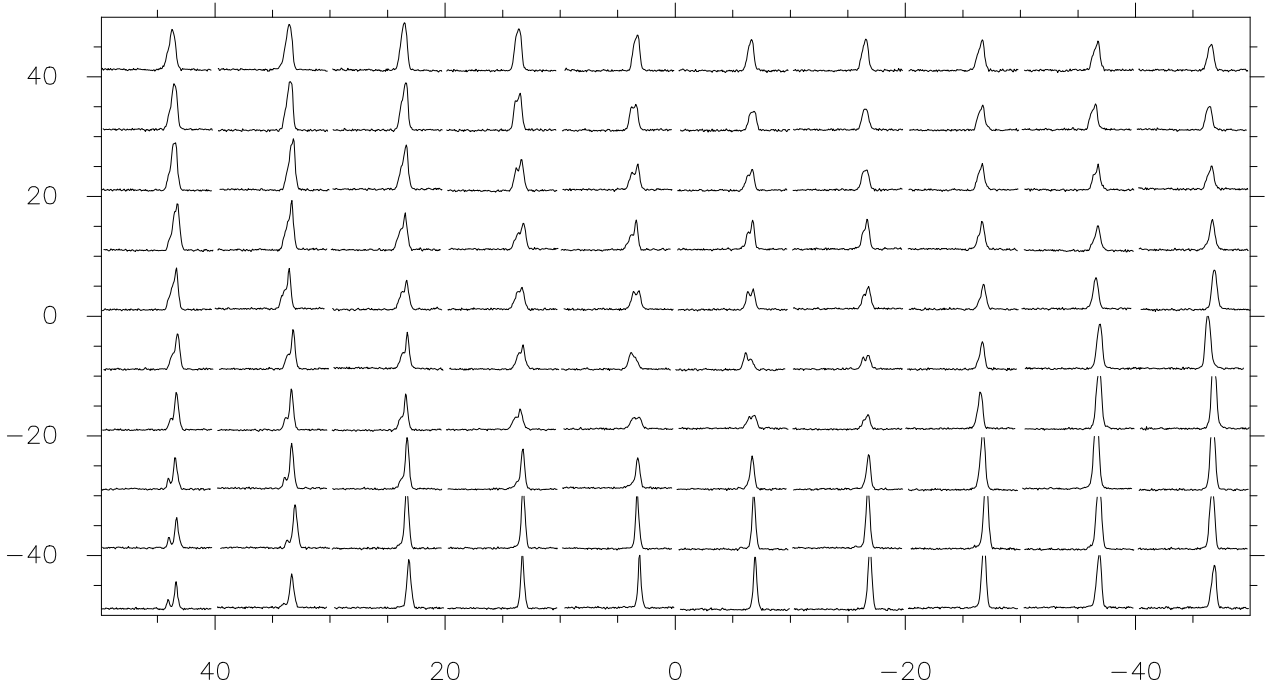}
\caption{$^{13}CO$(3-2) profiles observed over a 1\farcm 5 $\times$ 1\farcm 5 region around \object{IRAS\,10361-5830}. Each profile shows T$_{mb}$ in the interval from --1 to 25 K {\it vs.} LSR velocity in the range from --30 to --15 \kms. The (0,0) position coincides with the position of the IRAS source in J2000 equatorial coordinates.}
\label{fig:mapgrid13}
\end{figure*}

Figure \ref{fig:mapgrid13} displays the $^{13}$CO(3-2) spectra obtained for the observed region. The spatial separation between these profiles is 10\arcsec. Relative coordinates are expressed in arcseconds, referred to the IRAS source  position, centered at \radec\ = (10$^{h}$38$^{m}$0\fs 4, --58\degr 46\arcmin 17\farcs 8). The individual spectra exhibit two maxima with moderate emission and a separation of a few km s$^{-1}$ in the central part of the observed region, while the outer regions  generally show only one bright velocity component, except in the southeast section, where the two velocity components can be clearly identified. 

\begin{figure}
\centering
\includegraphics[angle=0,width=0.5\textwidth]{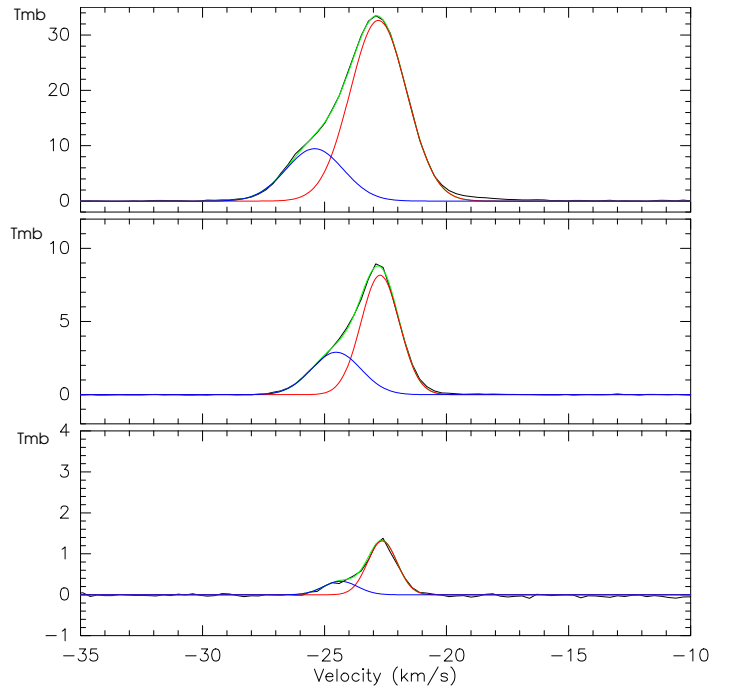}
\caption{$^{12}$CO(3-2), $^{13}$CO(3-2), and C$^{18}$O(3-2) averaged profiles in black lines. Gaussian fittings are overlaid in red and blue lines. The sum of the Gaussian components in each profile is shown in green.}
\label{fig:perfilespromedio2}
\end{figure}

Figure \ref{fig:perfilespromedio2} displays the $^{12}$CO(3-2), $^{13}$CO(3-2), and  C$^{18}$O(3-2) profiles (black lines) obtained by averaging the observed spectra.
In red and blue lines, we show the best Gaussian fitting to the profiles with two velocity components, and in green the sum of these.
This allows us to distinguish the different components in the observed area and get a global idea of the molecular gas parameters.
The figure shows the excellent correspondence of the Gaussian fitting with the averaged spectra, thus indicating the presence of gas at two different velocities in the central part of the observed region. The presence of two components at different velocities in the optically thin lines $^{13}$CO(3-2) and C$^{18}$O(3-2) suggests that the identification of the two components in the $^{12}$CO(3-2) spectrum is not caused by self-absorption effects.

The results are summarized in Table \ref{table:parametros}, which lists the number of Gaussian components, the peak velocity $\rm v$, the velocity width at half-intensity $\Delta {\rm v}$, the peak temperature $T_p$, and the integrated emission I ($=T_{p} \Delta$v). Values between brackets indicate the errors in the Gaussian parameters.

The emission is concentrated between --28 and --18 \kms, as estimated from the full width at 3$\sigma$ of the averaged profiles (10 \kms\ for the $^{12}$CO(3-2) spectrum, 7 \kms\  for $^{13}$CO, and 5 \kms\ for C$^{18}$O).   

\begin{table*}
\centering
\begin{tabular}{lccccccc}
\hline
{\it Line} & No. &  $\rm v$ [km s$^{-1}$] & $\Delta {\rm v}$ [km s$^{-1}$] & $T_p$ [K] & I [K km s$^{-1}$] & $T_{exc}$[K] & $\tau_{av}$   \\
\hline
$^{12}$CO(3-2) &  1  &  -25.71 (0.01)  &   2.29 (0.02)  &   8.33  &   20.3 (0.10)   & 20$\pm$1.3  & 10.5 \\
$^{12}$CO(3-2) &  2  &  -22.89 (0.01)  &   2.87 (0.01)  &  33.32  &  101.8 (0.17)   & 41$\pm$4    & 10.5 \\
               &     &                 &                &         &                 &             &      \\
$^{13}$CO(3-2) &  1  &  -24.35 (0.01)  &   2.63 (0.01)  &   3.05  &    8.6 (0.01)   & 20$\pm$1.3  & 0.3  \\
$^{13}$CO(3-2) &  2  &  -22.68 (0.01)  &   1.82 (0.01)  &   7.70  &   14.9 (0.01)   & 41$\pm$4    & 0.3  \\
               &     &                 &                &         &                 &             &      \\
C$^{18}$O(3-2) &  1  &  -24.34 (0.08)  &   1.62 (0.16)  &   0.33  &   0.56 (0.06)   &     ---     & ---  \\
C$^{18}$O(3-2) &  2  &  -22.63 (0.02)  &   1.41 (0.04)  &   1.31  &   1.97 (0.06)   &     ---     & ---  \\
               &     &                 &                &         &                 &             &      \\
HCO$^{+}$(4-3) &  1  &  -22.62 (0.08)  &   1.85 (0.24)  &   0.15  &   0.30 (0.03)   &     ---     & ---  \\
\hline
\end{tabular}
\caption{Parameters of the Gaussian fittings to the averaged spectra.}
\label{table:parametros}
\end{table*}

\begin{figure}
\centering
\includegraphics[width=0.5\textwidth]{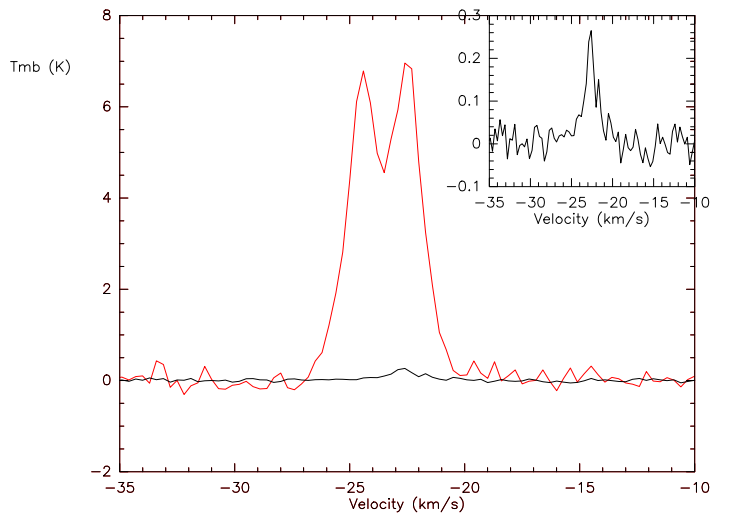}
\caption{$^{13}$CO(3-2) spectrum (red) overlaped with  HCO$^{+}$(4-3) spectrum (black) observed toward IRAS\,10361-5830.The inset show the HCO$^{+}$ spectrum at an appropriate scale.}
\label{fig:lineaspromedio}
\end{figure}

We adopt --25 \kms\ as the velocity of the faintest CO component and  --22.8 \kms\ as the velocity of the brightest one. It is clear from Fig. \ref{fig:mapgrid13} the remarkable decreasing in intensity of this last component towards the central region in comparison with the outer regions. 
The HCO$^{+}$ spectra shows only one component at --22.6 \kms, in coincidence with the brightest CO component. The HCO$^{+}$ spectrum is shown in Fig. \ref{fig:lineaspromedio} overlaid with the $^{13}$CO spectrum. No CS emission was detected. 

The detection of HCO$^{+}$ emission and the lack of CS emission indicates a region with densities of about 10$^{6}$ cm$^{-3}$, compatible with the critical density of HCO$^{+}$ ($n_{crit}= 1.8 \times 10^{6}$ cm$^{-3}$).

\begin{table*}
\centering
\caption{Infrared  sources and candidate YSOs projected onto the observed region. Sources 1, 2 and 3 correspond to sources 9, 20 and 21, respectively, in CNAV08.}
\begin{tabular}{clccccc}
\hline
{\it No.} & Source & RA, Dec.(J2000) & 2MASS counterpart & WISE counterpart & CLASS$^{1}$ \\
\hline
1 & \object{IRAS 10361-5830}        &  10$^{h}$38$^{m}$04\fs 00, -58\degr 46\arcmin 17\farcs 8 & --- & ---  & --- \\
2 & \object{MSX G286.3747-00.2630}  &  10$^{h}$38$^{m}$02\fs 18, -58\degr 46\arcmin 43\farcs 1 & 10380180-5846417 &J103801.52-584641.5  &  II \\
3 & \object{MSX G286.3773-00.2563}  &  10$^{h}$38$^{m}$04\fs 56, -58\degr 46\arcmin 24\farcs 8 & 10380461-5846233 &J103804.58-584620.9  &  II \\  
4 & --- & 10$^{h}$38$^{m}$05\fs08, -58\degr 45\arcmin 42\farcs 0 & --- &  J103805.8-584542 &  I \\
5 & --- & 10$^{h}$37$^{m}$58\fs04, -58\degr 46\arcmin 48\farcs 0 & --- &  J103758.4-584648 &  I \\
6 & --- & 10$^{h}$38$^{m}$00\fs07, -58\degr 46\arcmin 54\farcs 0 & --- &  J103800.7-584654 &  I \\
\hline
\end{tabular}
\label{table:fuentes}
\footnotesize $^{1}$From \citealt{Ohlendorf13}
\end{table*}

\subsection{Parameters of the molecular gas: excitation temperature and optical depth}

Excitation temperatures T$_{exc}$ and optical depths $\tau_{av}$ were estimated for the two velocity components  of the CO averaged spectra. $T_{exc}$ was evaluated from the optically thick ($\tau \gg$ 1) $^{12}$CO emission assuming Local Thermodynamic Equilibrium (LTE), using the expression:
\begin{equation}
T_{exc} = \frac{T^*} {ln\left[    \frac{T^*}{T^{12}_{p}+T^*J(T_{rad})^{-1}}   + 1 \right]},
\label{eq:texc2}                       
\end{equation}

\noindent where {\bf  $J(T)=(e^{\frac{T^{*}}{T}}-1)^{-1}$,}  $T^*$ = $\frac{h \nu_{32}}{k}$, $\nu_{32}$ is the frequency of the $^{12}$CO(3-2) line (345.79 GHz), $T^{12}_p$ is the peak temperature of the Gaussian in the $^{12}$CO spectrum, and $T_{rad}$ = 2.7 K. The error in  $T_{exc}$ was obtained adopting an uncertainty of 10\% in $T^{12}_p$.

Averaged optical depths $\tau^{12}_{av}$ and $\tau^{13}_{av}$, for $^{12}$CO and $^{13}$,CO respectively, were derived using the $^{13}$CO Gaussian fitting. To estimate $\tau^{13}_{av}$ we used the equation

\begin{equation}
\tau^{13}_{av} = - ln \left[  1 -  \frac{T^{13}_{p}}{T^{*}}  [ J(T_{exc})- J(T_{rad}) ]^{-1}  \right].
\label{eq:tau2}
\end{equation}

\noindent where the frequency of the $^{13}$CO(3-2) line was used in T$^{*}$.

Finally, the optical depth of the $^{12}$CO line was estimated from
\begin{equation}
\tau^{12}_{av}  = A  \tau^{13}_{av} \frac{\Delta \rm{v}^{13}}{\Delta \rm{v}^{12}} \left( \frac{\nu^{13}}{\nu^{12}}\right)^{2},
\label{eq:cocientetau}
\end{equation}

\noindent where $\Delta \rm{v}^{12}$ and $\Delta \rm{v}^{13}$ are the full width at half-maximum obtained from the averaged spectra, $\nu^{13}$ and ${\nu^{12}}$ are the frequency of the lines, and A = [$^{12}$CO]/[$^{13}$CO] = 70 is the isotopic ratio (\citealt*{LangerPenzias90},\citealt*{Wilsonrood94}). Results for $T_{exc}$ and $\tau_{av}$ are included in the last two columns of Table \ref{table:parametros}.

\begin{figure*}
\centering
\includegraphics[angle=0,width=0.9\textwidth]{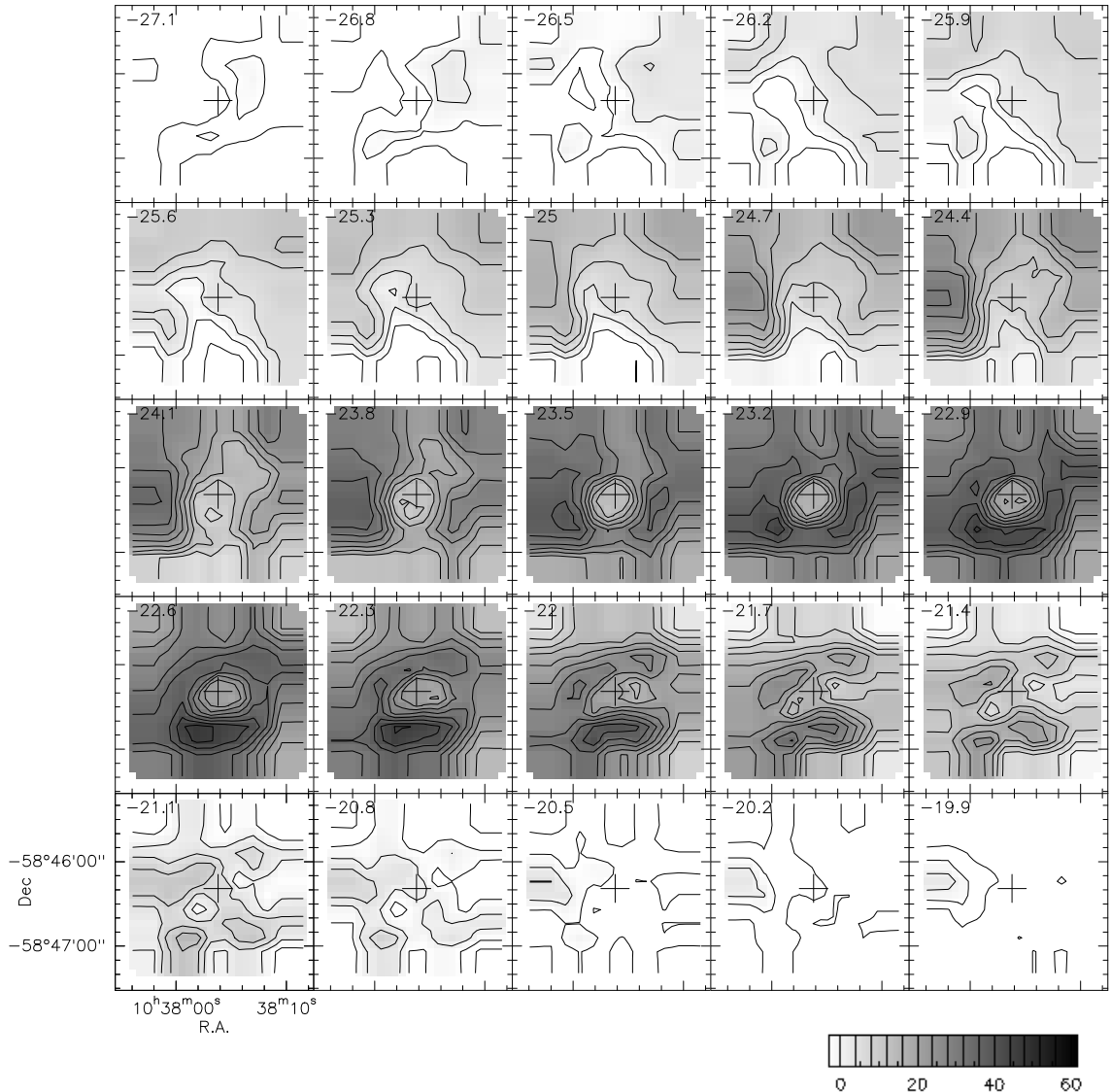}
\caption{Series of $^{12}$CO images showing the emission  in the range --27.1 to --19.9 \kms. The velocity of each panel is shown in the top-left corner of each panel. Contours go from 10 to 60 K in steps of 2 K. The central cross on each panel indicates the position of IRAS 10631-5830. The grey scale is indicated. Note that in this figure, R.A. increases towards the right.}
\label{fig:mapas12}
\end{figure*}

\subsection{Molecular gas distribution}

To study the spatial distribution of the molecular gas, we constructed a series of images of the $^{12}$CO emission at different velocities within the velocity interval from --27.1 to --19.9 \kms\  in steps of 0.3 \kms. These images, which are displayed in Fig. \ref{fig:mapas12}, show a low emission region centered at \radec\ = (10$^{h}$ 38$^{m}$ 04$^s$, --58\degr 46\arcsec 24\arcmin), close to the position of the IRAS source (indicated by a cross). The cavity is completely encircled by regions of enhanced emission, defining a molecular envelope. The low emission region displays its  steeper temperature gradient at {\rm v} = \hbox{--22.9$\pm$0.3} \kms.

 Towards more negative velocities, the cavity and the envelope can be clearly followed up to \hbox{--25.3} \kms, although faint extensions are present up to \hbox{--27.1} \kms, while towards more positive velocities, they extend up to \hbox{--20.8} \kms, with weaker extensions up to \hbox{--19.9} \kms.  
 
\begin{figure*}
\centering
\includegraphics[width=0.4\textwidth]{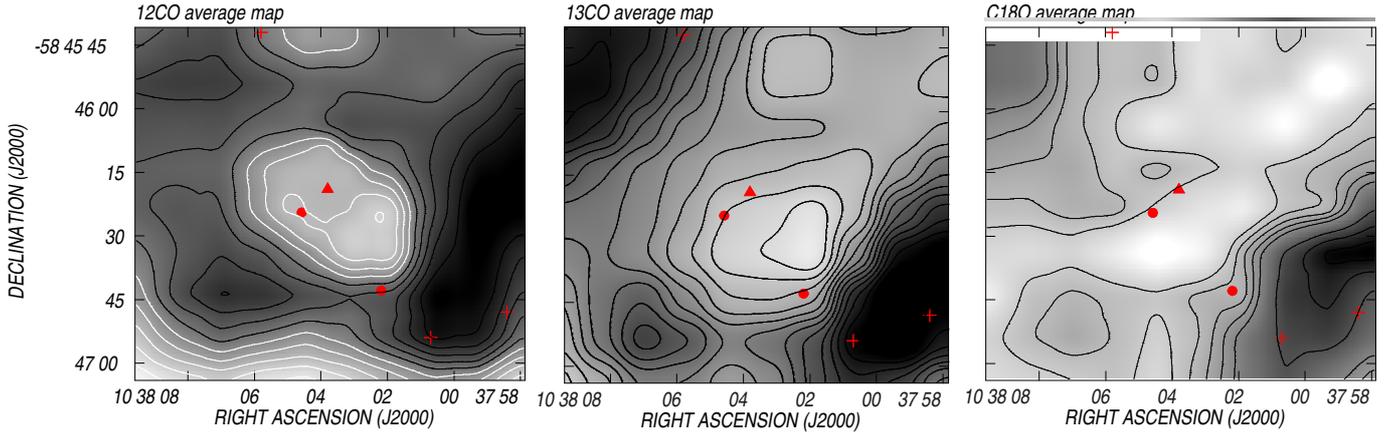}
\caption{{\it Left panel:} Average  $^{12}$CO line emission $T_{mb-12}$ within the velocity interval from --27.1 to --19.9 \kms. Contours go from 14 to 24 K, in steps of 1 K. {\it Middle panel:} Average image of the $^{13}$CO line emission showing $T_{mb-13}$ within the velocity interval from --27.1 to --20.8 \kms. Contours goes from 3 to 10 K in steps of 0.5 K. {\it Right panel:} Average image of the C$^{18}$O line emission showing $T_{mb-18}$ within the velocity interval from --25.0 to --21.1 \kms. Contours are 0.51, 0.68, 0.85, 1.02, 1.19, 1.70, 2.04, and 2.38 K.  The triangle indicates the position of IRAS 10361-5530. The different symbols mark the location of candidate YSOs identified by CNAV08 and \citet{Ohlendorf13}: MSX sources (circles) and WISE sources (plus signs). White contours in the left panel corresponds o T$_{mb} \le$ 17 K.}
\label{fig:promedios}
\end{figure*}

The structure can be imaged by averaging the $^{12}$CO emission within the velocity interval from --27.1 to --19.9 \kms. Figure \ref{fig:promedios} (left panel) displays the main-beam brightness-temperature T$_{mb-12}$ map of the $^{12}$CO(3-2) line emission. The position of the IRAS source and the candidate YSOs from CNVA08 and \citet{Ohlendorf13} are indicated in this figure. The coordinates and names of these sources are listed in Table \ref{table:fuentes}, along with the Class-type from \citet{Ohlendorf13}.

 The  $^{13}$CO(3-2) line emission distribution is shown in the  center panel of Fig. \ref{fig:promedios}, which displays the spatial distribution of $T_{mb-13}$ in the interval from --27.1 to --20.8 \kms. The right panel of the figure shows the C$^{18}$O emission in the interval --25.0 to --21.1 \kms.The {\it rms} noise of the averaged images is 0.07, 0.04, and 0.18 K for the $^{12}$CO, $^{13}$CO, C$^{18}$O images, respectively. 

The spatial distribution of the $^{12}$CO(3-2) line emission clearly shows the area of low intensity (with $T_{mb-12}$ $\lesssim$ 17 K) centered near the position of the IRAS source, encircled by a shell-like structure. The brightest region is located at \radec\ = \hbox{(10$^{h}$37$^{m}$59$^{s}$, --58\degr 46\arcmin 25\arcsec)}, and reaches about 24 K.
Very probably, this bright region corresponds to an extension towards the  West of \object{BYF\,77b} identified by \citet{Barnes11} in HCO$^{+}$(1-0) lines.
The envelope shows other areas of high emission located at \radec\ = \hbox{(10$^{h}$38$^{m}$8$^{s}$, --58\degr 45\arcmin 55\arcsec)}, coincident with \object{BYF\,77c} from the list of \citet{Barnes11}, and at \hbox{(10$^{h}$ 38$^{m}$ 07$^{s}$, --58\degr 46\arcmin 40\arcsec)}, which have intensities of about 20 K. In this map all the emission is above the {\it rms} noise ($3\sigma_{prom}=0.21$ K). 
The analysis of the  $^{13}$CO(3-2) line emission distribution (Fig.\ref{fig:promedios}, middle panel) leads to a similar conclusion, although in this case the envelope is not complete. Finally, denser regions in the molecular shell are detected in the C$^{18}$O line (right panel).


It is important to note that the velocity range where the $^{12}$CO, $^{13}$CO and C$^{18}$O emissions are detected is compatible with the velocity of the molecular gas associated with \object{Gum\,31} (CNAV08, \citealt{yonekura05}, \citealt{Barnes11}), indicating that the molecular shell identified with the APEX data is linked to \object{Gum\,31}. In particular, the velocity of BYF77 derived by \citet{Barnes11} coincides with the bright component at -22.8 \kms.

All candidate YSOs except source 3 (see Table \ref{table:fuentes}) coincide with molecular emission. Sources 2 and 4 are projected onto the borders of the molecular emission, while sources 5 and 6 appear projected onto regions of strong molecular emission. It is worth nothing that all Class I sources appear projected onto the molecular shell.

\subsection{Cold dust distribution}

\begin{figure*}
\centering
\includegraphics[width=0.9\textwidth]{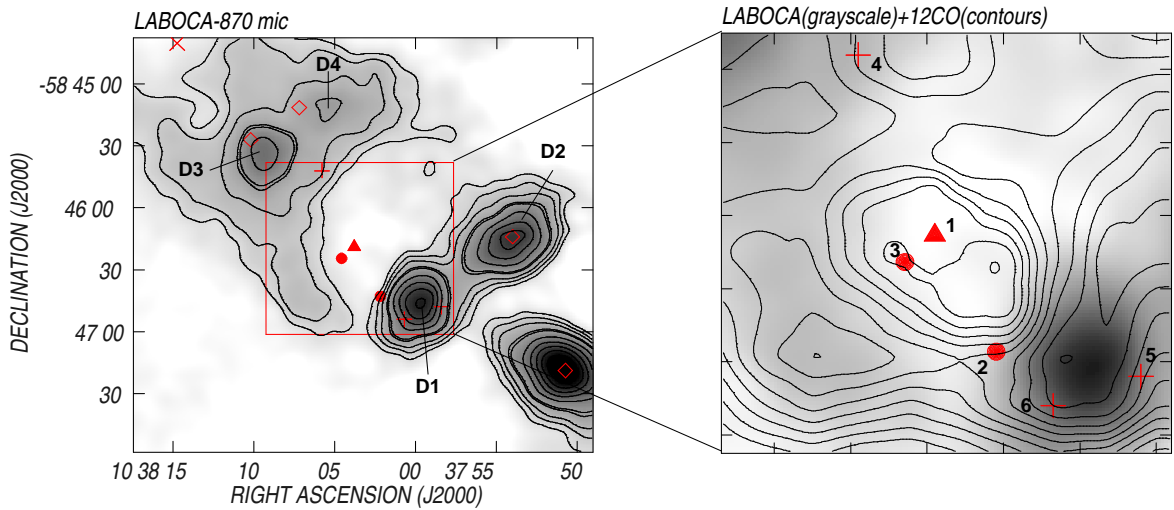}
\caption{{\it Left panel:} 870 $\mu$m continuum emission map. Dust clumps are indicated. The grayscale goes from -0.1 to 2.0 \jyb. Contour levels correspond to 0.2, 0.35, 0.5, and 0.7 \jyb, and from 0.75 to 2.0 \jyb\ in steps of 0.25 \jyb. The large square indicates the region observed in CO lines and the different symbols have the same meaning as in Fig.~\ref{fig:promedios}. {\it Right panel:} Overlay of the emission at 870 $\mu$m (grayscale) and the CO contours of Fig.~\ref{fig:promedios}. IR sources listed in Table \ref{table:fuentes} are indicated.}
\label{fig:laboca}
\end{figure*}

The continuum emission in the far-IR allowed us to investigate the distribution of the cold dust in the region. In the left panel of Fig.~\ref{fig:laboca}, we display the distribution of the emission at 870 $\mu$m, while in the right panel we show an overlay of this emission (in grayscale) and the $^{12}$CO emission (in contours) in the region observed in molecular lines, which is indicated with a red square in the left panel. The different symbols mark the position of candidate YSOs.

The image on the left shows dust cores centered at \radec\ = (10$^{h}$38$^{m}$00$^{s}$, --58\degr 46\arcmin 45\arcsec) (hereafter called D1), \radec\ = (\hbox{10$^{h}$37$^{m}$54$^{s}$}, \hbox{--58\degr 46\arcmin 15\arcsec}) (D2), \radec\ = (\hbox{10$^{h}$38$^{m}$10$^{s}$}, \hbox{--58\degr 45\arcmin 33\arcsec}) (D3), and \radec\ = (\hbox{10$^{h}$38$^{m}$05$^{s}$}, \hbox{--58\degr 45\arcmin 10\arcsec}) (D4). The last two cores appear embedded in extended emission. An additional core is detected at \radec\ = (\hbox{10$^{h}$37$^{m}$51$^{s}$}, \hbox{--58\degr 47\arcmin 18\arcsec}), quite further out of the region observed in molecular lines. The image on the right reveals the remarkable correspondence between the cold dust emission at 870 $\mu$m and the CO emission. In particular, core D1 coincides with the densest section of the molecular shell, seen in the C$^{18}$O image, and the extended dust emission to the west of the IRAS position agrees with the western section of the molecular shell. Moreover, the central cavity detected in CO lines is also present in the distribution of the continuum emission as a low emission region. These facts confirm the presence of a cold dust counterpart to the molecular shell. 
Finally, there is an excellent correspondence between the emission at 870 $\mu$m and the Herschel emission at 350 $\mu$m shown by \citet{Ohlendorf13}.

Source 4 is projected onto the border of the extended cold dust emission. Sources 5 and 6 coincide with D1, while source 2 appears projected onto its border. Additionally, J103810.2-584527 and J103807.2-584511 (indicated by diamonds), mentioned by \citet{Ohlendorf13}, are projected onto the borders of D3 and D4. The coincidence of these sources with both molecular gas and cold dust is compatible with their identification as YSOs.   

\section{Discussion}

\subsection{Comparison with infrared and radio continuum emissions}

\begin{figure*}
\centering
\includegraphics[width=0.9\textwidth]{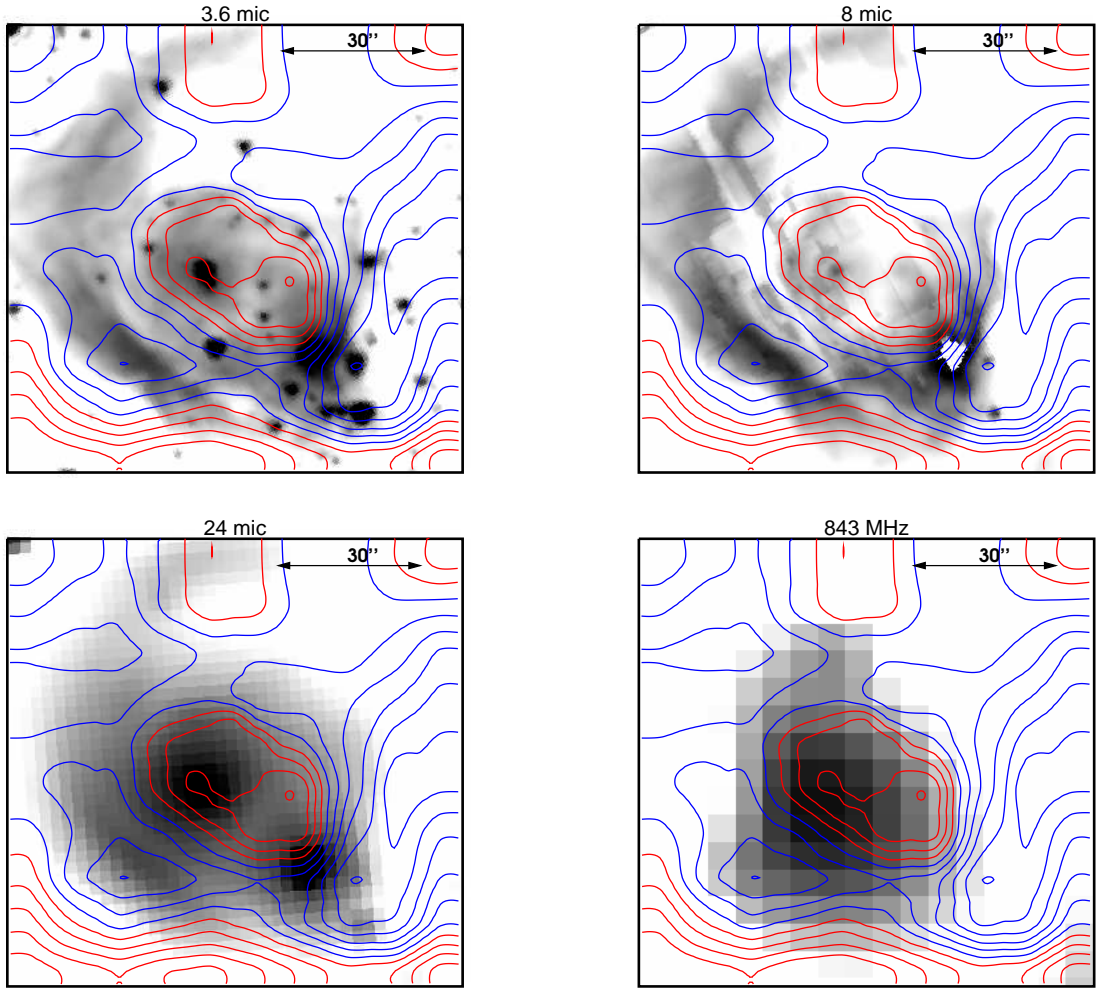}
\caption{$^{12}$CO(3-2) contours overlaid onto the IRAC emissions at 3.6 $\mu$m (top left panel) and 8 $\mu$m (top right panel), at 24 $\mu$m from MIPS (bottom left panel),  and  at 843 MHz from SUMSS (bottom right panel). The molecular cavity is indicated by the central red contours. }
\label{fig:co+8+24+843}
\end{figure*}

Figure \ref{fig:co+8+24+843} shows an overlay of the $^{12}$CO contours of Fig. \ref{fig:promedios} and the emissions at 3.6 $\mu$m from IRAC (top left panel), 8 $\mu$m from IRAC (top right panel), 24 $\mu$m from MIPS (bottom left panel), and 843 MHz from SUMSS (bottom right panel). 

In the 3.6 $\mu$m image we can see a number of point-like sources spread over the surveyed region, which were identified as belonging to a cluster by \citet{Ohlendorf13}. Source 3 (Class II) from Table \ref{table:fuentes} is projected near the center of the cavity detected in CO lines, and source 2 (Class II) appears close to the border of the cavity. Sources 4, 5 and 6 classified as Class I objects, are projected onto the dense molecular envelope. The arc-like structure mentioned by \citet{Ohlendorf13} is seen in coincidence with the denser regions of the molecular shell.

The emission at 8 $\mu$m arises mainly from both the polycyclic aromatic hydrocarbons (PAHs, \citealt{Leger84}) and from an underlying continuum emission attributed to very small grains, and is typical of photodissociation regions (PDRs). These molecules cannot survive inside \hii\ regions. They delineate the interface between ionized and molecular gas, indicating a clear interaction between them. 
Weak 8 $\mu$m emission partially coincident with the central cavity is probably located at this interface. The arc-like feature identified at 3.6 $\mu$m is also detected at 8 $\mu$m and at 5.8$\mu$m (not shown here and also mainly originated from PAH emission), and, as shown by \citet{Ohlendorf13}, can be identified in the MIPS emission at 24  $\mu$m, and in the Herschel image at 70  $\mu$m (see Fig. 12 from \citealt{Ohlendorf13}). This arc-like feature appears projected onto the bright CO emission regions, suggesting regions where photodissociation of the molecular gas is occurring.

The 24 $\mu$m emission arises from warm dust. In addition to a diffuse extended emission coincident with the cavity, the image shows that sources 2 and 3 (Class II) are the brighter sources in the mid-IR.

The continuum image at 843 MHz indicates the presence of radio emission within the shell. The origin of this emission is probably thermal and due to ionized gas, since it spatially coincides with emission at 24 $\mu$m caused by warm dust. The observed full width at half maximum of the source at 843 GHz is {\bf $45\arcsec\times 62\arcsec$}. The mean diameter of this source, after deconvolution with the synthesized beam of the telescope (HPBW= {\bf $43\arcsec\times 43\arcsec$} cosec($\delta$)), is $\leq$ 25\arcsec. The source is centered at \radec = (10$^{h}$38$^{m}$05\fs3, --58\degr46\arcmin27\farcs09), coincident within errors, with the position of source 3, suggesting the existence of ionized gas within the cavity and in the close environs of this source. A radio continuum image at higher frequency than 843 MHz with better angular resolution would be useful to investigate the ionized gas distribution within the cavity and to determine its physical parameters, since thermal emission is optically thick below 1 GHz.

The emission distribution at 8 and 24 $\mu$m, and in the radio continuum towards this region resembles that seen towards many infrared dust bubbles identified by \citet{Churchwell06} \citep[see][]{Watson09}, and is indicative of the action of UV photons and stellar winds in the environs of excitation sources. These bubbles appear also encircled by molecular gas (see for example \citealt{Zhang13}).

\subsection{Physical parameters of the molecular gas and dust}

As described in previous sections, our APEX data obtained with medium angular resolution revealed a molecular shell within clump 6 from the list of \citet{yonekura05} (see Fig. \ref{fig:grumo+fuentes}). In what follows we estimate its main physical parameters, which are summarized in Table \ref{parametrosregion}. 

We defined $\Delta$v$_{total}$ as the velocity range where the shell can be identified. It was estimated as $\Delta$v$_{total}$ = v$_{i}$-v$_{f}$, where v$_{i}$ y v$_{f}$ are the initial and final velocities at which the molecular emission is clearly detected in connection to this shell. These values were estimated from Fig. \ref{fig:mapas12} as v$_{i} = -20.8$ and  v$_{f} = -25.3$  km s$^{-1}$. 

The radius of the cavity $r_{cav}$ was estimated by taking into account the contour corresponding to $T_{mb-12}$ = 17 K as the inner radius of the envelope in the left panel of Fig. \ref{fig:promedios}. The result is given in arcseconds and pc. The outer radius of the envelope $r_{max}$ was measured following the outer contour of 17 K. Finally, the radius of the shell $r_{shell}$ was estimated as a mean value between $r_{cav}$ and $r_{max}$. 

The mean column density of $^{13}$CO, $N_{13CO}$, in the observed region was calculated  assuming LTE, using the equations of \citet*{Rolhfs04} and the image in the middle panel of Fig. \ref{fig:promedios}.  For the $^{13}$CO(3-2) we obtain,
\begin{equation}
\left[\frac{N_{13CO}}{cm^{-2}}\right] = C T_{exc} \int \tau^{13}(\nu)d\nu \left[ \frac{e^{\frac{2h\nu}{kT_{exc}}}}{e^{\frac{h \nu}{k T_{exc}}}-1} \right],
\label{eq:N13}
\end{equation}

\noindent where
\begin{equation}
 C = \frac{3}{7} \frac{16 \pi k \nu^2}{h c^3 A_{32}} = 7.9\times10^{13} 
\end{equation}

Folowing \citet*{Rolhfs04} we can approximate
\begin{equation}
T_{exc} \int \tau^{13}(\nu)d\nu \sim \frac{\tau^{13}}{1-e^{-\tau^{13}}} \int T_{mb}d{\rm v}.
\end{equation}

 As pointed out by \citet*{Rolhfs04}, this expression helps to eliminate to some extend optical depth effects. We adopted the optical depth of the $^{13}$CO(3-2) line corresponding to the most intense velocity component (see Table 2).
Finally, the average $H_{2}$ column density $N_{H2}= 1.1\times 10^{22}$ cm$^{-2}$ was obtained adopting $N_{H2}=7.7\times 10^{5} N_{13CO}$ \citep{Wilsonrood94}.
 
To calculate the total molecular mass $M_{H_{2}}$ in the observed region we applied
\begin{equation}
\left[\frac{M_{H_{2}}}{M_{\odot}}\right] = C' \left[\frac{N_{H_{2}}}{cm^{-2}}\right]  \left[\frac{Area}{\arcsec}\right] \left[\frac{d^{2}}{kpc}\right],
\label{eq:masa}
\end{equation}

\noindent where the constant $C' = 5.17 \times 10^{-25}$ includes the value of the solar mass 2$\times$10$^{33}$ g, the mean molecular weight $\mu$ = 2.76, derived after allowance of a relative helium abundance of 25\% by mass \citep{Allen73}, and the hydrogen atomic mass $m_H$ = 1.67$\times$10$^{-24}$ g. In this expression $Area$ is the area of the molecular emission within a region of $90\arcsec \times 90\arcsec$ in, and $d = 2.5\pm0.5$ kpc. We obtained $M_{H_{2}} = 290\pm 110 M_{\odot}$.
This value includes all the gas in the line of sight towards the observed region, which is $\sim$ 10-15\% of the total area of the clump.

Then, we derived the ambient density $n_{H_{2}}$ in the clump as
\begin{equation}
n_{H_{2}} = \frac{M_{H_2}}{\mu l^{2}L}
\end{equation}
We distributed the total molecular mass within the volume of a box of cross section equal to that of the observed area ($l^{2}$ with $l = 90\arcsec = 1.1$pc), and a size of $4\farcm0$ or 2.9 pc ($L$) in the line of sight, which corresponds to the size of clump 6 in the line of sight equal to its minor axis, as evaluated in Fig. \ref{fig:grumo+fuentes}. 


We estimated $n_{H_{2}}$$\sim$ 1200 cm$^{-3}$.  This value, is compatible with the ambient density obtained by Yonekura et al. (2005) (1600 cm$^{-3}$) based on C$^{18}$O(1-0) line data.

A rough estimate of the mass in the shell can be derived from the image in the middle panel of  Fig. \ref{fig:promedios}. Adopting the contour line corresponding to $T_{mb}^{13}$ = 4 K, as  the border of the central cavity  and using the same equations by \citet*{Rolhfs04} we estimate a molecular mass of 40$\pm$8 M$_{\odot}$.
The volume density in the shell, n$_{H2-shell}$, was estimated by distributing the molecular mass within the volume of a ring with inner and outer radii $r_{cav}$ and $r_{max}$, respectively. This value turns out to be 2300 cm$^{-3}$.

Following \citet{bohlinetal77}, the visual absorption in the observed region can be estimated assuming that gas and dust are well mixed from 
\begin{equation}
N_{HI}+2N_{H_{2}} = 2.5 \times 10^{21} A_{\rm v}
\end{equation}

\noindent Neglecting the \hi\ column density $N_{HI}$ and adopting $N_{H2}$=1.1$\times$10$^{22}$ cm$^{-2}$, a visual absorption $A_v$ = 9 mag is derived, compatible with $A_v$-values estimated by \citet{Preibisch14}.

From the LABOCA data, we determine the dust emission at 870 $\mu$m $I_{870}$, and we estimate the H$_2$ column density towards the (sub)mm cores (see Fig. \ref{fig:laboca}) using
\begin{equation}
\label{ircd}
N_{dust+gas}  = \frac{I_{870}}{B_{870}(T_d)\mu_{H_2} m_{\rm H} \kappa_{870} R_{\rm d}},
\end{equation}

\noindent where $I_{870}$= $S_{870}/\Omega_{beam}$, $\Omega_{beam}$ is the beam solid angle, $\kappa_{870}=1.0$ \cmdos g$^{-1}$ is the dust opacity per unit mass estimated for dust grains with thin ice mantles in cold clumps \citep*{Ossenkopf94}, and R$_{\rm d}$ is the adopted dust-to-gas ratio (= 1/100). The core total mass (gas + dust mass) can be obtained from
\begin{equation}
\label{irmass}
M_{dust+gas}  = \frac{S_{870} d^2}{B_{870}(T_d) \kappa_{870} R_{\rm d}}.
\end{equation}

Equations \ref{ircd} and \ref{irmass} where obtained from \citet*{Miettinen10}.



Table \ref{parametros870} summarizes the main parameters of the dust cores. Columns 1 to 3 identify the core and its coordinates, col. 4 lists the flux density at 870 $\mu$m, cols. 5 and 6, the total column density and  mass, respectively. Column 7 gives the effective radius of the core, and col. 8, the H$_{2}$ volume density. Derived densities are roughly compatible with the H$_{2}$ densities estimated from the molecular data. Note, however, that molecular data are available only for D1.

\begin{table}
\centering
\begin{tabular}{cc}
\hline
Parameter &  Value \\
\hline
$d$                     &  2.5$\pm$0.5 kpc \\
$\Delta {\rm v_{total}}$  &  6.3 km s$^{-1}$ \\
$T_{exc}$                &  41 $\pm$ 4 K \\ 
$\tau^{13}_{prom}$        &  0.3 $\pm$ 0.04 \\
$\tau^{12}_{prom} $       &  10.5 $\pm$ 1.05 \\
$r_{cav}$                &  18\farcs1 $\pm$ 2\farcs4 $\sim$ 0.22 $\pm$ 0.04 pc\\
$r_{max}$                &  35\farcs0 $\pm$ 7\farcs0 $\sim$ 0.42 $\pm$ 0.10 pc \\
$r_{shell}$              &  27\farcs0 $\pm$ 4\farcs9 $\sim$ 0.32 $\pm$ 0.06 pc\\
$M_{H_{2}-shell}$          &  40$\pm$8 $M_{\odot}$ \\
$n_{H_2-shell}$           &  $\sim$ 2300 cm$^{-3}$ \\
\hline
\end{tabular}
\caption{Parameters of the molecular shell.}
\label{parametrosregion}
\end{table}

\begin{table*}
\begin{tabular}{lccccccc}
\hline
              & $\alpha$(J2000) & $\delta$(J2000)                     & S$_{870}$             &   N$_{\rm dust+gas}$  &  M$_{\rm dust+gas}$   &  $r_{\rm eff}$  &  n$_{\rm H_2}$     \\

               & (h m s)         & ($^{\circ}$ ' ")                    &  Jy                   & 10$^{22}$cm$^{-2}$    &     $M_{\odot}$         &        pc       &   10$^3$ cm$^{-3}$    \\
\hline
D1     &   10 37 59.990  & --58 46 43.55        &  1.45 $\pm$  0.15              &    6.80 $\pm$ 1.05              &     55 $\pm$ 28                  &   0.30 $\pm$    0.05       &   6.40 $\pm$ 3.25 \\
D2     &   10 37 54.107  & --58 46 14.37        &   1.75 $\pm$ 0.18                &    6.15 $\pm$ 0.95             &     65  $\pm$ 36               &   0.35 $\pm$  0.07           &   5.52 $\pm$ 2.80 \\
D3      &  10 38 09.392 & --58 45 32.22         &  1.25 $\pm$ 0.12                &    4.25 $\pm$  0.65            &     46  $\pm$   26              &   0.35 $\pm$  0.07         &   3.50 $\pm$ 1.75 \\
D4      &  10 38 05.240 & --58 45 12.04         &  0.52 $\pm$  0.05               &    2.80 $\pm$ 0.42              &     20   $\pm$  11             &   0.30 $\pm$  0.06        &   2.30 $\pm$  1.30 \\
\hline
\end{tabular} 
\caption[]{Derived parameters of the Far- IR cores (870 $\mu$m).}
\label{parametros870}
\end{table*}

\subsection{Proposed scenario}

The presence of ionized gas, warm dust, and an extended PDR along with the high density derived for the region, point towards the existence of an \hii\ region, in agreement with the classification of source 3 as CHII region.

In this context, we propose that the \hii\ region originated in the photodissociation and ionization of part of the dense molecular clump. This interpretation is compatible with the observed decrease in intensity of the component at  --22.8 km s$^{-1}$ in the inner section of the shell, and the fact that the only component detected in HCO$^+$ coincides with the CO component at --22.8 \kms, which represents the bulk of the molecular gas in the clump.

We wonder if the molecular shell is expanding. \hii\ regions expand in their vecinity due to the difference in pressure between the ionized gas and the neutral gas that surrounds it. 
The difference between the H$_2$ ambient volume density n$_{H2}$ ($\sim$1200 cm$^{-3}$) and the volume density in the shell n$_{H2-shell}$ ($\sim$2300 cm$^{-3}$) suggests that the ionized region has started its expansion phase.
However, an inspection of the position-velocity $^{12}$CO maps towards the cavity do not show clear signs of expansion. 
Moreover, bearing in mind an uncertainity of 20\% in distance, errors in derived volume densities may be as high as 140\%, and the difference between the two volume densities should be taken with caution.
A scenario compatible with our findings suggests that the \hii\ region is young, close to the end of its formation phase and/or just starting its expansion phase (see Dyson y Williams 1998).

\begin{figure}
\centering
\includegraphics[angle=270,width=0.5\textwidth]{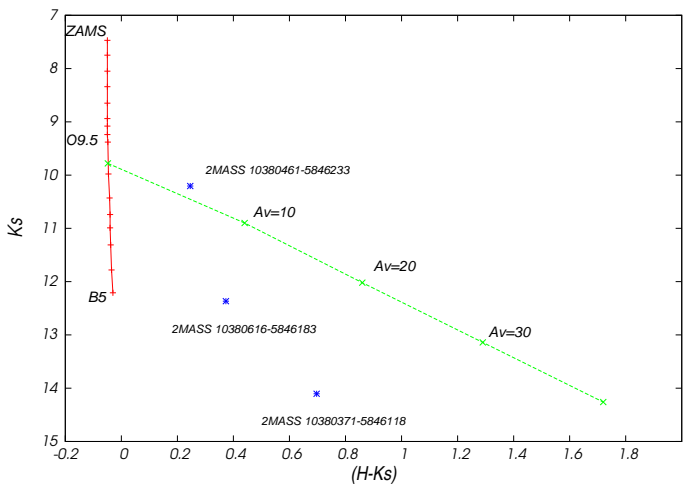}
\caption{CM diagram for the AAA-sources within the cavity. Absolute magnitudes for the ZAMS in the K$_{s}$ band were obtained from \citet{Koornneef83}. The  green dashed line shows the reddening vector.}
\label{fig:CM-CC}
\end{figure}

We have investigated the presence of exciting sources within the cavity using the 2MASS catalogue taking into account sources having  excellent photometric quality (AAA). The analysis of the sources was performed in a region of 20\arcsec\ in radius centered at \radec=(10${^h}$38$^{m}$04$^{s}$,\hbox{--58\degr 46\arcmin 20\arcsec}). The color-color (CC) (H-K$_{s}$,J-H)- and color magnitude (CM) (H-K$_{s}$,K$_{s}$)-diagrams (the CC diagram is not shown here) revealed that 2MASS 10380461-5846233, coincident with MSX\,G286.3773-00.2563, is a candidate O9V star (H-K$_{s}$ = 0.246, J-H = 0.505). The position of this source in the CM-diagram is indicated in Fig. \ref{fig:CM-CC}, along with those of the other two AAA-sources found in the region. The 2MASS source 10380616-5846183 is a candidate B star, while 10380371-5846118, would be a later-B star.

\citet{Ohlendorf13} estimated for 2MASS 10380461-5846233 a stellar mass of $\sim$5.8 M$_{\odot}$ from a SED obtained using \citeauthor{Robitaille07}'s (2007) tool, which is not compatible with a O9V star. However, their result should be taken with caution since the fitting tool has some limitations (\citealt{Deharveng12}, \citealt{Robitaille08}, \citealt{Offner12}).

\begin{figure}
\centering
\includegraphics[angle=270,width=0.5\textwidth]{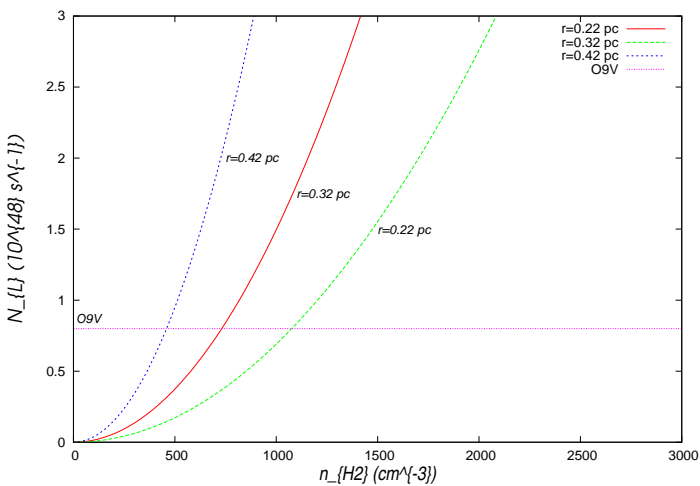}
\caption{UV photon flux N$_L$ {\it vs.} H$_2$ ambient density. The curves indicate the UV flux N$_L$ as function of n$_{H_2}$ corresponding to different Str$\ddot{o}$mgren radius. The horizontal line shows the theoretical UV flux for an O9V type star \citep{Smith02}.}
\label{fig:flujos}
\end{figure}

An \hii\ region developed within the cavity would have electron densities $n_e$ $\simeq$ 2$n_{H2}$ $\simeq$ 2.4$\times$10$^3$ cm$^{-3}$, taking into account that the original material is molecular hydrogen. These electron densities, along with the small radius of the cavity, are roughly compatible with the presence of a compact \hii\ region (r $\le$ 0.25 pc and $n_e \ge$ 5$\times$10$^3$ cm$^{-3}$, \citealt{Kurtz05}). The diagram of Fig. \ref{fig:flujos} shows the UV flux $N_{L}$ {\it vs.} $n_{H2}$. The curves indicate the UV photon flux necessary to ionize an \hii\ region as a function of the H$_2$ ambient density $n_{H2}$ = $n_e$/2 for Str$\ddot{o}$mgren radius of 0.22, 0.32 and 0.42 pc (corresponding to $r_{cav}$, $r_{shell}$ and $r_{max}$ respectively). The horizontal line marks the UV flux corresponding to an O9V-type star (see \citealt{Martins02}). According to the diagram, the UV flux $N_L$ of an O9V star would be enough to create an \hii\ region of 0.22 pc in size with a density of $\sim$ 1.2$\times$10$^{3}$ \cm. 

We can also speculate that the presence of stellar winds can modify this picture. Stellar winds sweep-up the surrounding material creating low density cavities encircled by expanding shells \citep{Weaver77}.
Thus, the stellar wind from the O9V star may sweep-up part of the ionized gas, thus decreasing the density of the \hii\ region and favoring the formation of an \hii\ region of 0.22-0.32 pc in radius. Additional studies are necessary to test this possibility.



 
\section{Summary}

Based on $^{12}$CO, $^{13}$CO, C$^{18}$O, HCO$^+$, and CS line observations and millimeter continuum at 870 $\mu$m obtained with the APEX telescope, we analyzed the spatial distribution and the main physical parameters of the molecular gas and the cold dust associated with a dense clump linked to IRAS\,10361-5830, located in the environs of the \hii\ region Gum\,31 at 2.5$\pm$0.5 kpc. 

The study allowed us to identify a molecular shell in the dense clump centered on the IRAS position. The analysis of the molecular emission reveals the existence of two velocity components projected onto the central cavity: a faint component centered at --25 \kms\ and a bright one centered at  --22.8 \kms, while generally only the bright velocity component appears projected onto the molecular shell. These velocities coincide with those of the material linked to Gum\,31. The cold dust counterpart of the molecular shell is detected at 870 $\mu$m, as well as in Herschel data at 350 $\mu$m.
  
A comparison of the spatial distribution of the molecular shell with the emission at 24 $\mu$m reveals the presence of warm dust inside the molecular shell, suggesting the existence of ionizing sources, while the radio continuum image at 843 MHz shows  diffuse emission coincident with the central molecular cavity and \object{MSX G286.3773-00.2563}. The IRAC-Spitzer emission at 8 $\mu$m shows the presence of an arc-like structure, also detected at 3.6, 5.8, and 24 $\mu$m, coincident with the densest region in the shell, suggesting the existence of a photodissociation region at the interface between the ionized and molecular material.

The molecular shell has a mean radius of 0.32 pc and is detected within a velocity interval of 6.3 \kms. We have estimated a molecular mass in the shell of 40$\pm$8 M$\odot$. 

A number of candidate YSOs classified as Class I and II objects appear projected onto the central cavity and the molecular shell. 
The presence of candidate YSOs indicates that star formation is active in this dense clump.
The 2MASS source 10380461-5846233 (MSX G286.3773-00.2563), candidate to late O-type star seems to be responsible for creating a compact \hii\ region inside the molecular clump.


\begin{acknowledgements}
We acknowledge the anonymous referee for useful comments.
This project was partially financed by CONICET of Argentina under projects PIP 02488 and PIP 00356, and UNLP under project 11/G120. M.R. is supported by CONICYT of Chile through grant No. 1080335.
This publication is based on data acquired with the Atacama Pathfinder Experiment (APEX). APEX is a collaboration between the Max-Planck-Institut fur Radioastronomie, the European Southern Observatory, and the Onsala Space Observatory.
This research has made use of the NASA/ IPAC Infrared Science Archive, which is operated by the Jet Propulsion Laboratory, California Institute of Technology, under contract with the National Aeronautics and Space Administration. This work is based [in part] on observations made with the Spitzer Space Telescope, which is operated by the Jet Propulsion Laboratory, California Institute of Technology under a contract with NASA. This publication makes use of data products from the Two Micron All Sky Survey, which is a joint project of the University of Massachusetts and the Infrared Processing and Analysis Center/California Institute of Technology, funded by the National Aeronautics and Space Administration and the National Science Foundation. The MSX mission is sponsored by the Ballistic Missile Defense Organization (BMDO).
\end{acknowledgements}

\bibliographystyle{aa}
\bibliography{paper}

\end{document}